\documentclass[intlimits,twoside,a4paper]{article}
\usepackage{euscript}
\usepackage[cp1251]{inputenc}

\usepackage[eqsecnum]{cmpj3}

\usepackage{bm}

\issue{2021}{24}{1}{13601}
\doinumber{10.5488/CMP.24.13601}

\title[Orientational instability of the director]%
{Orientational instability of the director in a nematic cell caused by 
electro-induced anchoring modification
}
\author[O.S. Tarnavskyy, M.F. Ledney] {O.S. Tarnavskyy, M.F. Ledney}
\address{Faculty of Physics, Taras Shevchenko National University of Kyiv, Hlushkova Avenue, 4, 03022, Kyiv, Ukraine
}

\date{Received June 16, 2020, in final form September 15, 2020}
\begin{document}

\maketitle

\begin{abstract}
We theoretically investigate the threshold for the director reorientation from the homeotropic state to the hybrid homeotropic-planar state and vice versa
in a cell filled with a flexoelectric nematic liquid crystal (NLC)
subjected to an electric field. 
The liquid crystal is doped by a CTAB-like substance,
a part of molecules of which dissociates into positive and negative ions.
The anchoring on one of the cell surfaces is assumed to be strong and homeotropic, while the other surface can adsorb positive ions which play the role of an orienting surfactant for NLC molecules on this surface.
At certain voltages, the orientational transitions in the bulk of the NLC are possible due to the changing conditions for the director on the adsorbing surface.
We calculate respective threshold voltages as functions of anchoring parameters. 
The existence of the critical values of these parameters,
beyond which the orientational transitions do not take place, is established.

\keywords nematic liquid crystal, orientational instability,
threshold of orientational instability, boundary conditions, flexopolarization 

\end{abstract}

\section{Introduction}
\label{intro}

In recent decades, the rapid development of liquid crystal display technologies has been stimulated by intensive studies in the physics of liquid crystals (LC).
An LC cell is the basic constructional element of any liquid crystal display.
A wide practical use of LCs is based on exploitation of their unique electro- and magneto-optical properties.
These properties are closely related to the orientational order of LC molecules.
One of the most well-known and broadly used LC orientational phenomena is the Fr\'{e}ede\-ricksz transition, i.e., the threshold reorientation of the nematic liquid crystal (NLC) director caused by external electric or magnetic fields~\cite{deGennes_book}.
In spite of the fact that the threshold reorientation of the director is a bulk phenomenon, its characteristics, such as the threshold fields and director reorientation extent, substantially depend on the conditions for the director on the cell surfaces. 
These conditions are determined by the anchoring energy, the easy axis orientation, etc.
Usually, it is not possible to change the anchoring characteristics dynamically since they are established during the cell fabrication~\cite{Wu_book}. 
Therefore, orientational transitions in LC cells predominantly occur at fixed anchoring parameters.
Nevertheless, as it is shown in~\cite{Nazarenko_1994, Pinkevich_1996}, a spontaneous Fr\'{e}edericksz transition induced by changes in conditions for the director on the cell surfaces is possible.
Hence, the possibility to dynamically change  the anchoring parameters  is very attractive since it allows one to control the boundary conditions for the director on the cell surfaces, which could be used to improve the characteristics of the existing electrooptical devices or to create the new ones.

As recent studies suggest, under some conditions the director easy axis induced on the polymer substrate of a cell can change its orientation under the action of light or low-frequency electric/magnetic fields~\cite{Kurioz2002, Janossy2004, Ant2011, Lee2011, Pasechnik2013, Joly2004};
this, in turn, affects the equilibrium director configuration in the bulk of the NLC.
One of the mechanisms that can be used to adjust the type and strength of the anchoring is ion adsorption on the cell surfaces~\cite{Barbero_book}.   
Thus, the addition of the dopant $\mathrm{CTAB}$ molecules, which dissociate producing ions~$\mathrm{CTA}^+$ and~$\mathrm{Br}^-$, to the NLC host,
can induce the homeotropic orientation of the director on the cell surface capable of adsorbing positive ions~$\mathrm{CTA}^+$~\cite{Proust_1972, Kulkarni_2016}.
At a sufficient surface density, these adsorbed ions can orient NLC molecules homeotropically in the layer adjacent to the surface, using their long elastic tails, independently of the way NLC molecules were oriented in the absence of adsorption.
This mechanism was used in a number of experiments~\cite{Sutormin_2012,Sutormin_2013,Sutormin_2014,Sutormin_2016,Sutormin_2016_2},
where, owing to adsorption/desorption phenomena, the possibility to dynamically control the anchoring parameters  using an electric field was demonstrated.
In particular, the authors of~\cite{Sutormin_2012,Sutormin_2016} studied the dynamics of orientational transitions from the homeotropic NLC director configuration to the homeotropic-planar configuration and vice versa
after the field was turned on and subsequently turned off.
These transitions had thresholds, although the dependence of the threshold voltages on NLC cell parameters was not considered. 

In this paper, we theoretically study the threshold orientational instability of the director in a flexoelectric NLC cell subjected to an electric field that is created by a constant electric potential difference between the cell surfaces.
The NLC is doped by a CTAB-like substance, a part of the molecules of which dissociates in the NLC medium into positive and negative ions.
The NLC anchoring with one of the surfaces is assumed to be homeotropic and strong.
The other surface can adsorb positive ions and provides the planar anchoring in the absence of adsorbed ions.
We assume that the anchoring parameters, namely, the strength and the type, i.e., whether the anchoring is homeotropic or planar, depend on the density of adsorbed positive ions. 

We consider orientational transitions between stable director configurations in the bulk of the NLC.
These transitions are caused by the changes in the type and strength of the anchoring induced by variations in the applied voltage.
We calculate the threshold voltages of the orientational transitions from the homeotropic configuration to the homeotropic-planar configuration of the director and vice versa;
the dependence of these voltages on the parameters of the homeotropic and planar anchoring is studied.
We find the admissible ranges of the anchoring parameters within which the orientational transitions are possible.
The paper is organized as follows.
In section~\ref{sec:2}, the free energy of the NLC cell along with the equations for the director and the electric field is presented.
In section~\ref{sec:3}, the spatial profile of the electrostatic potential in the bulk of the NLC is calculated at the values of parameters that correspond to the vicinity of the orientational transition threshold.
In section~\ref{sec:4}, the threshold voltages are calculated.
The dependence of the threshold voltages on parameters of the homeotropic and planar anchoring is discussed in section~\ref{sec:5}.
Our conclusions are presented in section~\ref{sec:6}.

\section{Free energy of the NLC and equations for the director and the electric field}
\label{sec:2}

We consider an NLC cell restricted by planes~$z=0$ and~$z=L$.
Anchoring between the NLC director and the upper surface,~$z=L$ is assumed to be strong.
The director orientation on this surface is homeotropic, i.e., the director is perpendicular to the surface at each of its points.
The director easy axis on the lower surface~$z=0$ is directed along this surface, namely, along the $Ox$-axis; so, the anchoring on this surface is planar.
The NLC is doped by a $\mathrm{CTAB}$-like substance, a part of the molecules of which dissociates in the NLC medium into positive and negative ions.
We assume that the surface~$z=L$ does not adsorb ions, while the surface~$z=0$ is capable of adsorbing positive ions.
At a relatively small surface density of adsorbed ions, the anchoring on the lower surface remains planar.
However, as this density increases, the planar anchoring on the lower surface can vanish completely, and 
homeotropic anchoring induced by adsorbed ions emerges instead~\cite{Barbero_1994}.
The constant potential difference~$U$ is maintained between cell surfaces, which creates the static electric field~$\bf E$ in the bulk of the NLC directed along the $Oz$-axis. 
This field considerably affects the spatial distribution of charges in the bulk as well as the density of charges adsorbed on the lower surface.
As mentioned above, a variation in this adsorption density changes the type and strength of the NLC anchoring with the adsorbing surface.
Thus, we try to answer the question how the voltage~$U$ applied to the cell provokes the orientational instability of the director in the bulk.
Note that orientational transitions induced by an electric field in such NLC cells were observed in~\cite{Sutormin_2012, Sutormin_2014, Sutormin_2016}.
The free energy of the NLC cell can be written as follows:
\begin{equation}\label{energy}
F=  F_\text{el}+F_E+F_{S},
\end{equation}
\begin{equation*}
F_\text{el}=  \dfrac{1}{2} \int\limits_V \Bigl\{ K_1
({\rm div}\, {\bf n})^2+K_2({\bf n}\cdot{\rm rot}\,{\bf n})^2 +K_3[{\bf n}\times{\rm
rot}\, {\bf n}]^2\Bigr\}\, \rd V, 
\end{equation*}
\begin{equation*}
F_E=  \int\limits_V \left(-\dfrac{1}{8\piup}{\bf E}\hat\epsilon{\bf E}-{\bf P}{\bf E}+\rho \varphi\right)\, \rd V, 
\end{equation*}
\begin{equation*}
F_{S}= -\dfrac{1}{2}\int\limits_S \!\!\left[W_h ({\bf n}{\bf e}_z)^2+W_p ({\bf n}{\bf e}_x)^2\right]\rd S,
\; W_h>0,\, W_p>0.
\end{equation*}
Here,~$F_\text{el}$ is the elastic energy of the NLC, $F_E$ accounts for the anisotropic and flexoelectric contributions made by an electric field into the free energy of the NLC as well as the contribution made by ions~\cite{Ponti_1999}, $F_{S}$ has the form of the Rapini potential and describes the anchoring between the NLC and the surface of the lower surface, $K_1$, $K_2$, $K_3$ are elastic constants, $\bf n$ is the director,
$\hat\epsilon=\epsilon_\perp\hat{\bf 1}+\epsilon_a{\bf n}\otimes {\bf n}$,
$\epsilon_a=\epsilon_\parallel-\epsilon_\perp>0$ are
the static electric permittivity tensor of the NLC and its anisotropy, ${\bf E}$, $\varphi$ are the electric field and its potential,
$\rho$ is the bulk density of charges,
${\bf e}_x$, ${\bf e}_z$ are unit vectors of the Cartesian coordinate system,
$W_h$, $W_p$ are anchoring energies related to the deviation of the director on the surface~$z=0$ in homeotropic and planar directions, respectively,
${\bf P}=e_1{\bf n}\,{\rm div}\,{\bf n}-e_3[{\bf n}\times{\rm rot}\,{\bf n}]$, $e_1$, $e_3$ are
the flexoelectric polarization and flexoelectric coefficients.

The surface free energy~$F_S$ is taken in a simple form that generalizes the Rapini model~\cite{Rapini_1969, Zhao_2002} on the basis of the following considerations.
The anchoring between the NLC and the lower surface is planar in the absence of adsorbed positive ions.
This planar anchoring gradually weakens while the homeotropic anchoring, which is created by adsorbed ions, emerges with an increasing surface density of adsorbed ions~$\sigma$.
To give a quantitative description of interaction between the NLC and the adsorbing surface, we assume that energies of the homeotropic~$W_h$ and planar~$W_p$ anchoring depend on the surface density of adsorbed ions.
Thus, we take the homeotropic anchoring energy in the form that generalizes the one proposed in~\cite{Barbero_1994},
\begin{equation}\label{Wh}
W_h=W_{0h}\dfrac{\sigma}{\sigma_m}\left(1-\dfrac{\sigma}{\sigma_m}\right),
\end{equation}
where $W_{0h}$ is entirely determined by properties of the substrate material and by the way the substrate surface has been treated ($W_{0h}>0$),
$\sigma_m$ is the largest possible surface density of adsorbed ions (the maximum number of vacancies on the surface).
As it follows from (\ref{Wh}), the homeotropic anchoring is the strongest if adsorbed ions occupy a half of all vacancies, $\sigma=0.5\sigma_m$~\cite{Barbero_1994}; we assume that the planar anchoring is absent under this condition.
It is obvious that the electric field created by the potential difference~$U$ applied between the cell surfaces stimulates the transfer of positive ions from the adsorbing surface into the bulk of the NLC.
A decrease in the surface density~$\sigma$ of adsorbed ions causes not only quantitative but also qualitative changes in the anchoring between the NLC and the adsorbing surface.
Thus, in this case, the homeotropic anchoring weakens while the planar anchoring, on the contrary, becomes stronger. 
Taking the above-mentioned features into account, we write the planar anchoring energy in the form
\begin{equation}\label{Wp}
W_p=\left\{ \begin{array}{lr}
W_{0p}\left(1-2\sigma/\sigma_m\right), & \quad \mbox{if}\quad \sigma\leqslant 0.5\sigma_m, \\
0, & \quad \mbox{if}\quad \sigma> 0.5\sigma_m,
\end{array}\right.
\end{equation}
where $W_{0p}$ depends on the way the surface has been treated and on the properties of its material ($W_{0p}>0$).

\begin{figure*}[!t]
\centering
\hspace{8mm}
\includegraphics[width=62mm]{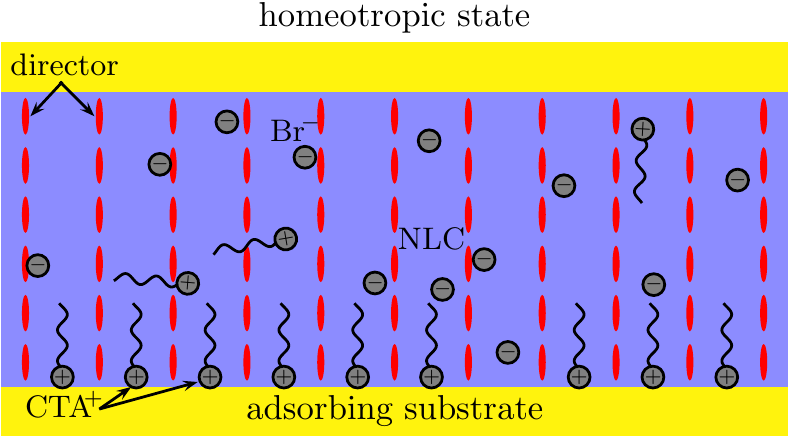}
\hspace{2mm}
\includegraphics[width=68mm]{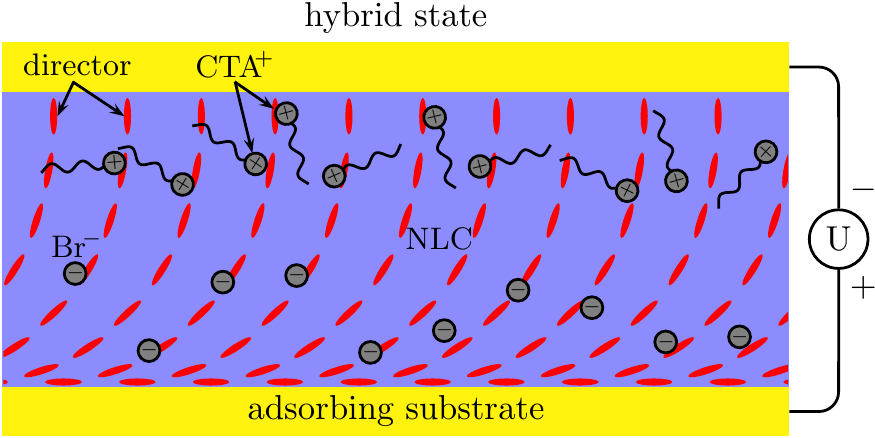}\\[0.2cm]
\hspace*{3mm}(a)\hspace{62mm}(b)
\caption{(Colour online) Schematic picture of homeotropic~(a) and hybrid~(b) director configurations in the cell.}
\label{cell}
\end{figure*}

We assume that in the absence of the external electric field~($U=0$),
the adsorption density~$\sigma$ is sufficient to ensure the homeotropic anchoring on the lower cell surface.
Thus, the director in the bulk of the cell is oriented homogeneously along the $Oz$-axis (see figure~\ref{cell}a).
As it follows form these considerations, at a certain threshold voltage~$U$ the orientational transition in the bulk of the NLC can occur owing to the changes in boundary conditions for the director on the adsorbing surface.
It means that the initial homeotropic orientation of the director becomes unstable and turns into the hybrid homeotropic-planar orientation (see figure~\ref{cell}b).

We consider planar deformations of the director, so the hybrid director reorientation takes place in the $xOz$-plane.
Owing to homogeneity of the system along the $Oy$-axis, the director in the bulk of the NLC can be written as follows:
\begin{equation}\label{director}
{\mathbf n}=\bigl(\sin\theta(z),0,\cos\theta(z)\bigr),
\end{equation}
where~$\theta$ is an angle that the director makes with the $Oz$-axis.

By minimising the free energy~(\ref{energy}) with respect to the angle $\theta$ and taking into account the expression for the director~(\ref{director}), we arrive at the following stationary equation
\begin{equation}\label{bulkequ}
(K_1\sin^2\theta+K_3\cos^2\theta)\theta''_{zz} -
\left((K_3-K_1) \theta'^2_z + e\varphi''_{zz}+\dfrac{\epsilon_a \varphi'^2_z}{4\piup}\right) \sin\theta\cos\theta = 0
\end{equation}
and respective boundary conditions
\begin{equation}
\begin{split}\label{boundary1}
& \!\!\!\!\!\! \left[(K_1\sin^2\theta+K_3\cos^2\theta)\theta'_z - (e\varphi'_z + W_h - W_p)\dfrac{\sin 2\theta}{2}\right]_{z=0} \!\!\!\!\!\!=0, \\
& \!\!\!\!\!\! \left.\theta\right|_{z=L}=0.
 \end{split}
\end{equation}
Here, $e=e_1+e_3$; $W_h$ and $W_p$ are defined by (\ref{Wh}) and~(\ref{Wp}).
The prime symbol denotes derivatives with respect to~$z$.

It is obvious that equation~(\ref{bulkequ}) for the angle~$\theta$ should be solved together with electrostatics equations for the electric field~$\bf E$.
By minimising the free energy~(\ref{energy}) with respect to the potential~$\varphi$, we obtain
\begin{equation}\label{potentialequ}
\dfrac{\rd}{\rd z}\left[-(\epsilon_\perp+\epsilon_a\cos^2\theta)\varphi'_z-2\piup e\theta'_z\sin 2\theta
\right] = 4\piup\rho.
\end{equation}
Boundary conditions for equation~(\ref{potentialequ}) can be written as
\begin{equation}\label{potentialboundary}
\left.\varphi\right|_{z=0}=0,\qquad \left.\varphi\right|_{z=L}=-U.
\end{equation}

The bulk density of charges in the right hand side of equation~(\ref{potentialequ}) equals $\rho=q(n_+-n_-)$, where $q$ is the elementary charge.
Assuming that the bulk density of ions can be described by the Boltzmann distribution, we define the bulk densities of positive~$n_+$ and negative~$n_-$ ions as it is done in~\cite{Barbero_2004}
\begin{equation}\label{rho_charge}
n_+(z)=\dfrac{n_0}{Z_+} \exp \left(-\frac{q\varphi(z)}{k_\text{B}T}\right), \quad
n_-(z)=\dfrac{n_0}{Z_-} \exp \left(\frac{q\varphi(z)}{k_\text{B}T}\right),
\end{equation}
where~$n_0$ is the bulk density of CTAB molecules which have dissociated into ions, $k_\text{B}$ is the Boltzmann constant, $T$ is temperature.
The surface density of positive ions adsorbed on the lower surface is given by the expression~\cite{Barbero_2004}
\begin{equation}\label{sigma_charge}
\sigma=\dfrac{N_0}{Z_+} \exp\left(\frac{A}{k_\text{B}T}\right),
\end{equation}
where $A$ is the adsorption energy, $N_0=n_0L$.
The values of $Z_+$ and~$Z_-$ can be found from the condition that the numbers of positive as well as negative ions in the system per unit area of the surface is equal to~$N_0$
\begin{equation}\label{statints}
Z_+=\re^\frac{A}{k_\text{B}T}+\dfrac{1}{L}\int\limits_0^L\re^{-\frac{q\varphi(z)}{k_\text{B}T}}\,\rd z,\qquad
Z_-=\dfrac{1}{L}\int\limits_0^L\re^{\frac{q\varphi(z)}{k_\text{B}T}}\,\rd z.
\end{equation}

\section{Potential profile in the bulk of the NLC}
\label{sec:3}

We are interested in the threshold voltage at which the initial homogeneous director configuration in the bulk of the cell becomes unstable.
In the vicinity of the threshold, the angles describing the director deviations are small ($|\theta|\ll1$).
Thus, we can linearize equation~(\ref{bulkequ}) and boundary conditions~(\ref{boundary1}) in~$\theta$
\begin{equation}\label{thetaequ}
K_3\theta''_{zz}-\left(e\varphi''_{zz}+\dfrac{\epsilon_a\varphi'^2_z}{4\piup}\right)\theta=0,
\end{equation}
\begin{equation}\label{boundary1lin}
\bigl[-K_3\theta'_z + (e\varphi'_z + W_h - W_p)\theta\bigr]_{z=0}=0, \qquad
\theta\bigl|_{z=L}=0
\end{equation}
as well as the equation for the potential~(\ref{potentialequ})
\begin{equation}\label{potentialequlin}
\varphi''_{zz} = -\dfrac{4\piup}{\epsilon_{||}}\rho
\end{equation}
and boundary conditions~(\ref{potentialboundary}).
As can be seen, in the linear approximation in~$\theta$, the equation for the potential~(\ref{potentialequlin}) does not depend on the angle~$\theta$.
This allows us to find the spatial profile of the potential~$\varphi(z)$.
By introducing a dimensionless coordinate~$\zeta=\dfrac{z}{L}$,
a potential~$u=\dfrac{q\varphi}{k_\text{B}T}$, a voltage~$v=\dfrac{qU}{k_\text{B}T}$, an adsorption energy~$\alpha=\dfrac{A}{k_\text{B}T}$ and a square of the dimensionless screening parameter~$\delta=\dfrac{4\piup q^2 N_0 L}{\epsilon_{||}k_\text{B}T}$, we can rewrite~(\ref{potentialequlin})
with boundary conditions~(\ref{potentialboundary}), taking into account~(\ref{sigma_charge}) and~(\ref{statints}),
as follows:
\begin{equation}\label{dimpotequ}
u''_{\zeta\zeta}=-\delta\left(\dfrac{\re^{-u}}{Z_+}-\dfrac{\re^{u}}{Z_-}\right),
\end{equation}
\begin{equation}\label{dimpotentialboundary}
u(0)=0,\qquad u(1)=-v,
\end{equation}
where
\begin{equation}\label{ZZsigma}
Z_+=\re^\alpha+\int\limits^1_0 \re^{-u(\zeta)} \rd\zeta,\;
Z_-=\int\limits^1_0 \re^{u(\zeta)} \rd\zeta, \; \sigma=N_0\dfrac{\re^\alpha}{Z_+}.
\end{equation}
To find the solution~$u(\zeta)$ of equation~(\ref{dimpotequ}), which satisfies the boundary conditions~(\ref{dimpotentialboundary}),
we use methods of the calculus of variations.
To this end, we first note that equation~(\ref{dimpotequ}) can be obtained by minimization of the functional
\begin{equation}\label{functional}
S[u]=\dfrac{1}{2\delta}\int\limits^1_0 u'^2_\zeta \,\rd\zeta + {\rm ln}\Big(\re^\alpha+
\int\limits^1_0 \re^{-u(\zeta)}\,\rd\zeta\Big) + {\rm ln}\Big(\int\limits^1_0 \re^{u(\zeta)}\,\rd\zeta\Big),
\end{equation}
where the function~$u(\zeta)$ satisfies~(\ref{dimpotentialboundary}).
\begin{figure}[t]
\centering
\includegraphics[width=65mm]{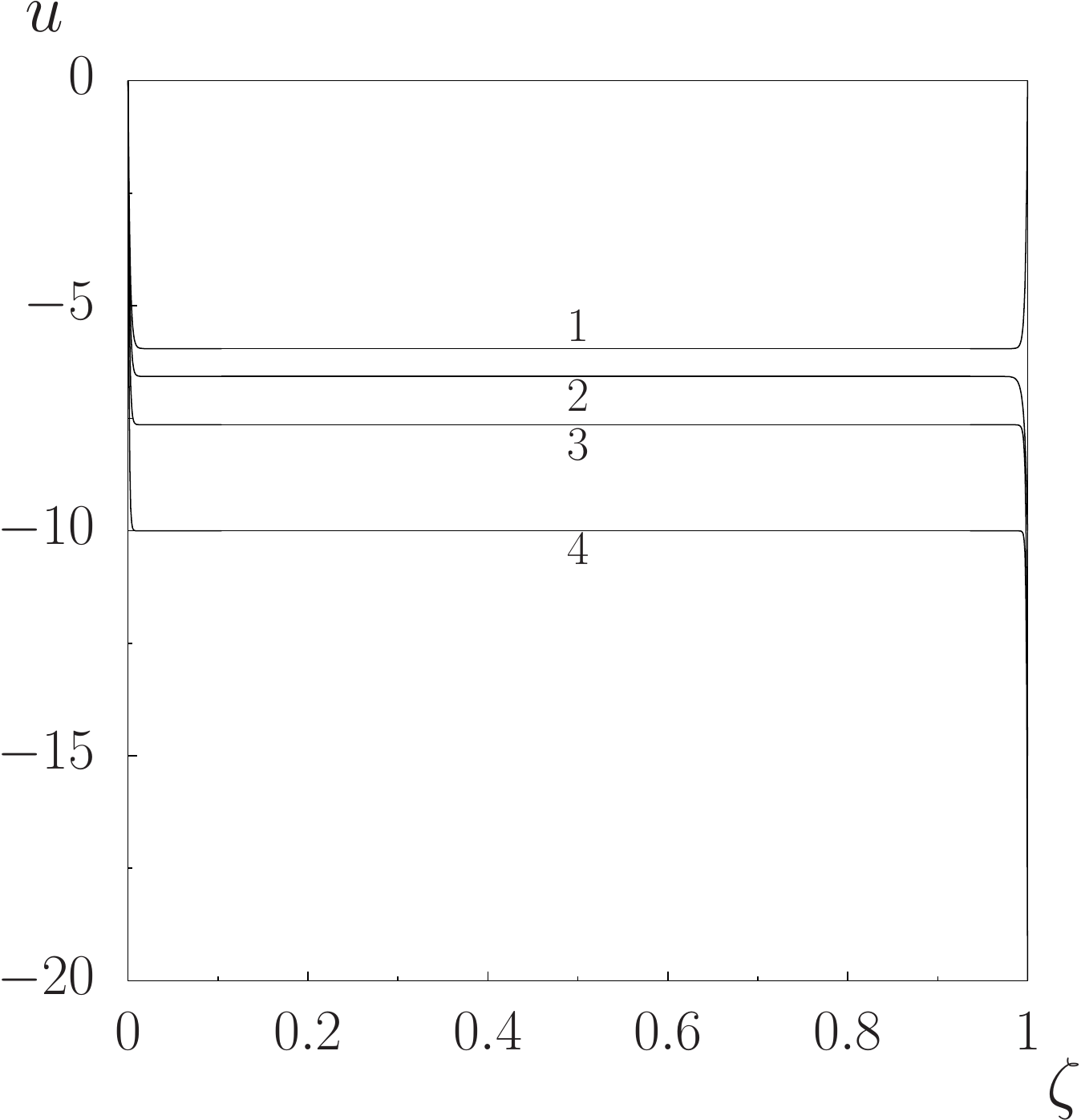}
\caption{Dimensionless electric potential $u$ versus $\zeta$ calculated for several values of the dimensionless applied voltage $v$ at the values of other parameters presented in table~\ref{vartable}. $v=0$ (1), $10$ (2), $15$ (3), $20$ (4).}
\label{potentialfig}
\end{figure}

The substitution of expression~(\ref{ansatz}) for the potential $u(\zeta)$ into (\ref{ZZsigma}) yields the value of~$\sigma$ at the given voltage~$v$
\begin{equation}\label{sigma}
\sigma\!=\!\dfrac{N_0 \re^\alpha}{\re^\alpha+\int\limits^1_0 \re^{-u(\zeta)}\,\rd\zeta}\!=\!
\dfrac{N_0}{1+\re^{\beta-\alpha}[1+\lambda f(-\beta)+\mu f(v-\beta)]}.
\end{equation}

\begin{figure}[t]
\centering
\includegraphics[width=65mm]{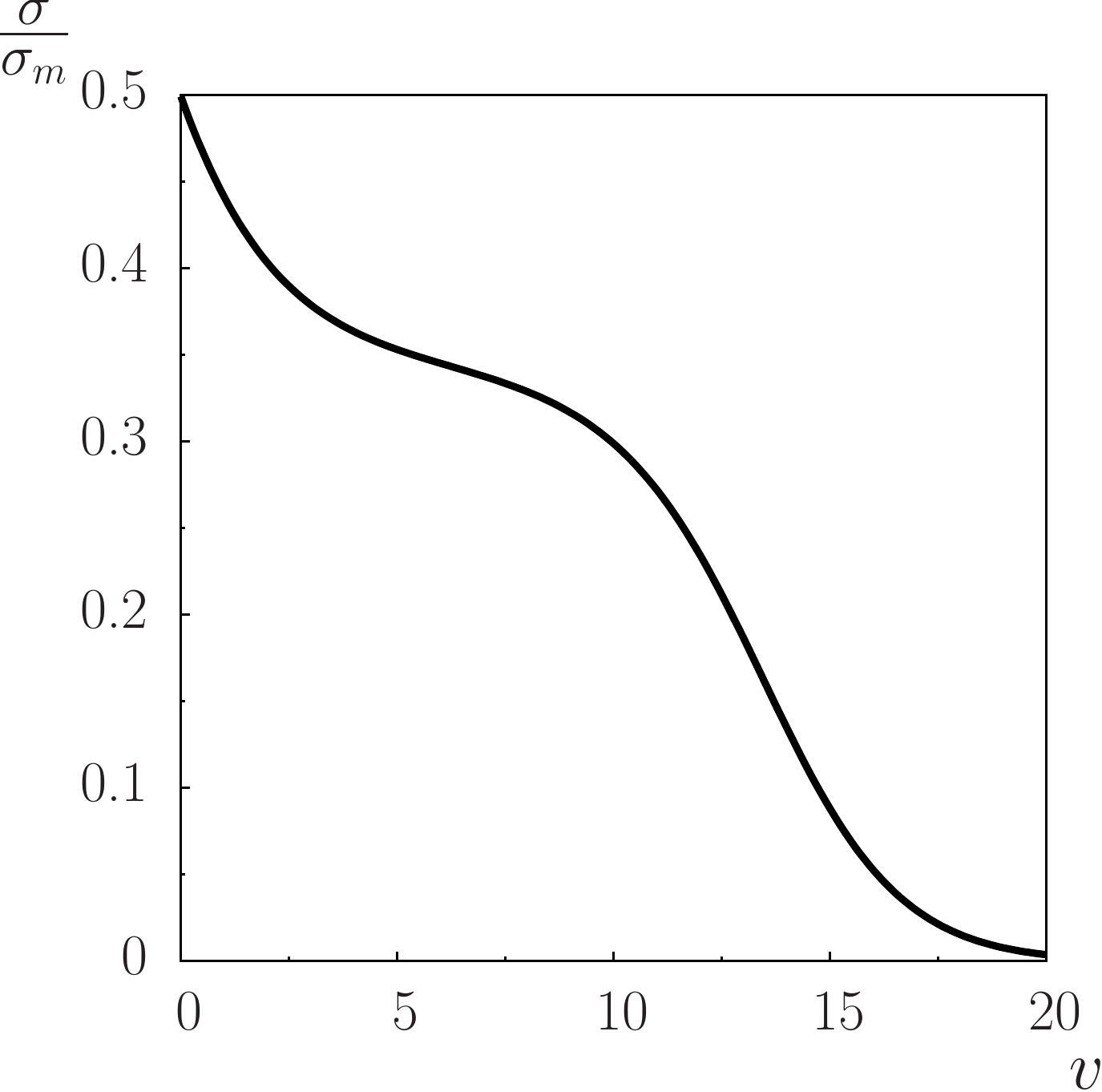}
\caption{Surface density of adsorbed ions on the lower cell surface versus the dimensionless applied voltage~$v$ at the values of parameters presented in table~\ref{vartable}.}
\label{sigmafig}
\end{figure}

Let the bulk density~$n_0$ of CTAB molecules that have dissociated in the NLC be sufficient to ensure that~$\delta\gg1$.
Hence, at relatively small applied voltages~$v$, a complete screening of the electric field in the bulk of the liquid crystal takes place. 
The spatial profile of the electrostatic potential in this case can be described with a sufficient accuracy by the expression
\begin{equation}\label{ansatz}
u(\zeta)=\beta \re^{-\zeta/\lambda}-(v-\beta)\re^{-(1-\zeta)/\mu}-\beta,
\end{equation}
where~$\beta$, $\lambda$ and $\mu$ are unknown constants ($\lambda\ll 1$, $\mu\ll 1$) that should be found by minimising the functional~$S[u]$ with respect to them.
As can be seen, the expression for the potential~(\ref{ansatz}) satisfies the boundary conditions~(\ref{dimpotentialboundary}) with the accuracy up to the terms of the order of~$\re^{-1/\lambda}$ and~$\re^{-1/\mu}$.
By substituting the potential~$u(\zeta)$~(\ref{ansatz}) into the functional~(\ref{functional}) and performing integration (see Appendix for details) with the accuracy up to the terms of the order of $\re^{-1/\lambda}$ and $\re^{-1/\mu}$, we obtain the following function of three variables~$\beta$, $\lambda$ and~$\mu$
\begin{equation}\label{S}
\begin{split}
S(\beta,\lambda,\mu) & =  \dfrac{\beta^2}{4\delta\lambda}+\dfrac{(v-\beta)^2}{4\delta\mu}+
{\rm ln}\bigl(\re^{\alpha-\beta}+1+\lambda f(-\beta)
\\
&
+\mu f(v-\beta)\bigr) +{\rm ln} \bigl(1+\lambda f(\beta)+\mu f(\beta-v)\bigr),
\end{split}
\end{equation}
where
\begin{equation}\label{f}
f(x)=\sum\limits^\infty_{k=1}\dfrac{x^k}{k\cdot k!}
\end{equation}
are rapidly convergent power series.
By minimising the function~$S(\beta,\lambda,\mu)$ (\ref{S}) with respect to all its arguments, we find the values of~$\beta$, $\lambda$ and $\mu$.
Substituting the latter values into~(\ref{ansatz}), we obtain the spatial profile of the potential~$u(\zeta)$ in the cell.
In figure~\ref{potentialfig}, the potential~$u(\zeta)$ calculated for the values of parameters close to typical~\cite{Sutormin_2012, Sutormin_2014, Sutormin_2016, Barbero_book, Blinov_book} (see table~\ref{vartable}) is presented.
As can be seen, at these values of parameters, the case of the electric field screening inside the NLC cell takes place. 

\begin{table}[htb]
\caption{Values of parameters used in calculations.}
\label{vartable}
\vspace{2ex}
\begin{center}
\begin{tabular}{lcc}
\hline
\noalign{\smallskip}
Parameter & Denotation & Value \\
\noalign{\smallskip}\hline\noalign{\smallskip}
Temperature & $T$ & $300$\,K\\
Cell thickness & $L$ & $10$~\textmu m\\
Adsorption energy & $A$ & $5\,k_\text{B}T$ \\
\raisebox{-0.5em}{\parbox{3.5cm}{\vspace{3pt}
Number of ions per unit surface\vspace{3pt}}} &
$N_0$ & $10^{17}\, \text{m}^{-2}$\\
\raisebox{-0.5em}{\parbox{3.5cm}{\vspace{3pt}
The largest possible\\[0pt] density of adsorbed ions\vspace{3pt}}} &
$\sigma_m$ & $0.562\cdot 10^{17}\, \text{m}^{-2}$\\
NLC elastic constant & $K_3$ & $10\,\text{pN}$ \\
\raisebox{-0.75em}{\parbox{3.5cm}{\vspace{3pt}
Eigenvalues of static\\[0pt] electric permittivity\\ tensor\vspace{3pt}}}& \raisebox{-0.7em}{\parbox{1cm}{\centering $\epsilon_{||}$\\[5pt] $\epsilon_\perp$}} & \raisebox{-0.7em}{\parbox{1cm}{\centering $19.7$\\[5pt]$6.7$}}\\
\raisebox{-0.3em}{\parbox{3.5cm}{\vspace{3pt}
Sum of flexoelectric\\[0pt] coefficients}} & $e_1+e_3$ & $0$ or $20\, \text{pC/m}$\\
\noalign{\smallskip}\hline
\end{tabular}
\end{center}
\end{table}

In figure~\ref{sigmafig}, we present the surface density~$\sigma$ of adsorbed ions on the lower surface as a function of the potential difference~$v$ between the cell surfaces, which was calculated using formula~(\ref{sigma}).
As the voltage~$v$ increases, the surface density~$\sigma$ drops quite rapidly and approaches zero at voltages of the order of~0.5~V.

\section{Threshold voltages}
\label{sec:4}

We next rewrite the linearized equation~(\ref{thetaequ}) for the director angle~$\theta$ and respective boundary conditions~(\ref{boundary1lin}) as follows
\begin{equation}\label{thetaequlindimensionless}
\theta''_{\zeta\zeta}-(\nu\varkappa u''_{\zeta\zeta}+\varkappa^2u'^2_\zeta)\theta=0,
\end{equation}
\begin{equation}\label{boundary1lindimensionless}
\begin{split}
\! & -\left.\dfrac{\theta'_\zeta}{\theta}\right|_{\zeta=0}\!\!\!\!\!\!-\nu\dfrac{\varkappa\beta}{\lambda}+
\varepsilon_\text{h}\dfrac{\sigma}{\sigma_m}\left(1-\dfrac{\sigma}{\sigma_m}\right)-\varepsilon_\text{p}
\left(1-2\dfrac{\sigma}{\sigma_m}\right)=0, \\
\! & \theta\bigr|_{\zeta=1}=0.
\end{split}
\end{equation}
Here, we use dimensionless temperature~$\varkappa=\dfrac{k_\text{B}T}{q}\sqrt{\dfrac{\epsilon_a}{4\piup K_3}}$,
flexoelectric parameter~$\nu = (e_1+e_3)\sqrt{\dfrac{4\piup}{\epsilon_a K_3}}$ and parameters of the homeotropic~$\varepsilon_\text{h}=\dfrac{W_{0h} L}{K_3}$ and planar~$\varepsilon_\text{p}=\dfrac{W_{0p} L}{K_3}$ anchoring.

The solution of equation~(\ref{thetaequlindimensionless}) on the interval~$0\leqslant \zeta \leqslant 1$ can be found taking into account an explicit form~(\ref{ansatz}) of the potential~$u(\zeta)$.
We make use of the fact that some terms in our problem are exponentially small.
Hence, in the interval~$0\leqslant\zeta\leqslant 1/2$ with the accuracy up to exponentially small terms, we have~$u'_\zeta\approx-\dfrac{\beta}{\lambda}\re^{-\zeta/\lambda}$,
$u''_{\zeta\zeta}\approx\dfrac{\beta}{\lambda^2}\re^{-\zeta/\lambda}$.
Introducing a new variable~$t=\varkappa\beta \re^{-\zeta/\lambda}$, we rewrite equation~(\ref{thetaequlindimensionless}) in the form
\begin{equation}\label{tequ}
t^2\theta''_{tt}+t\theta'_t-(\nu t+t^2)\theta=0.
\end{equation}

In the interval~$1/2 \leqslant\zeta\leqslant 1$, taking into account approximate relations
$$
u'_\zeta\approx\frac{\beta-v}{\mu}\re^{-(1-\zeta)/\mu} \text{ and }
u''_{\zeta\zeta}\approx\frac{\beta-v}{\mu^2}\re^{-(1-\zeta)/\mu}
$$
and introducing a new variable $t=\varkappa(\beta-v)\re^{-(1-\zeta)/\mu}$, we can rewrite equation~(\ref{thetaequlindimensionless}) in the same form~(\ref{tequ}).

It is worth noting that in the absence of the flexoelectric polarization ($\nu=0$), equation~(\ref{tequ}) is a modified Bessel equation~\cite{Abramowitz_book}.
However, if the NLC possesses flexoelectric properties~($\nu\ne 0$), then the solutions of equation~(\ref{tequ}) cannot be easily expressed in terms of special functions.
Therefore, we seek one of the solutions in the form of power series~$X(t)=\sum\nolimits^\infty_{n=0} a_n t^n$, where~$a_n$ are unknown coefficients.
Substituting~$X(t)$ into~(\ref{tequ}) and equating the coefficients of respective powers of~$t$, we obtain the following recurrence relation for~$a_n$
\begin{equation}\label{a-n}
a_n=\dfrac{\nu a_{n-1}+a_{n-2}}{n^2}, \quad n\geqslant 2,
\end{equation}
$a_1=\nu$ and~$a_0=1$.

As is known from the theory of ordinary differential equations~\cite{Zwillinger_book}, the other linear independent solution of equation~(\ref{tequ}) can be found in the form $X(t){\rm ln}|t|+Y(t)$, where $Y(t)=\sum\nolimits^\infty_{n=0} b_n t^n$ and $b_n$ are unknown coefficients.
The substitution of this solution into equation~(\ref{tequ}) yields a recurrence relation for coefficients $b_n$
\begin{equation}\label{b-n}
b_n=\dfrac{\nu b_{n-1}+b_{n-2}-2n a_n}{n^2},\quad n\geqslant 2,
\end{equation}
$b_1=-\nu$ and~$b_0=1$.

Therefore, the solutions of equation~(\ref{tequ}) in upper and lower halves of the cell read
\begin{equation}\label{theta1}
\begin{split}
\theta_1(\zeta) & \!=\!\left(\!A\!+\!B\,{\rm ln} \bigl|\varkappa\beta \re^{-\frac{\zeta}{\lambda}}\bigr| \right)\,
\!\!X\bigl(\varkappa\beta \re^{-\frac{\zeta}{\lambda}}\bigr)\!+\!
B\,Y\bigl(\varkappa\beta \re^{-\frac{\zeta}{\lambda}}\bigr),\\
& \mbox{if} \quad 0\leqslant\zeta\leqslant 1/2, \\
\theta_2(\zeta) & \!=\!\left(\!C\!+\!D\,{\rm ln} \bigl|\varkappa(\beta-v)\re^{-\frac{(1-\zeta)}{\mu}}\bigr| \right)\,
\!\!X\bigl(\varkappa(\beta\!-\!v)\re^{-\frac{(1-\zeta)}{\mu}} \bigr)\\
&+D\,Y\bigl(\varkappa(\beta\!-\!v)\re^{-\frac{(1-\zeta)}{\mu}} \bigr), \quad \mbox{if} \quad 1/2\leqslant\zeta\leqslant 1,
\end{split}
\end{equation}
where~$A$, $B$, $C$ and $D$ are unknown coefficients.
Using the fact that the function $\theta(\zeta)$ and its derivative~$\theta'(\zeta)$ are continuous at~$\zeta=1/2$, we have
\begin{equation}\label{tseam1}
\theta_1\left(\varkappa\beta \re^{-\frac{1}{2\lambda}}\right)=\theta_2\left[\varkappa(\beta-v) \re^{-\frac{1}{2\mu}}\right],
\end{equation}
\begin{equation}\label{tseam2}
-\frac{\varkappa\beta \re^{-\frac{1}{2\lambda}}}{\lambda}\theta'_{1\zeta}\!\left(\!\varkappa\beta
\re^{-\frac{1}{2\lambda}}\!\right)\!=\frac{\varkappa(\beta-v)\re^{-\frac{1}{2\mu}}}{\mu}
\theta'_{2\zeta}\!\left[\!\varkappa(\beta-v) \re^{-\frac{1}{2\mu}}\!\!\right]\!\!.
\end{equation}
From equations~(\ref{tseam1}) and~(\ref{tseam2}), taking into account the boundary condition~$\theta(\zeta=1)=0$~(\ref{boundary1lindimensionless}) which takes on the form~$\theta_2\bigl[\varkappa(\beta - v)\bigr]=0$ and neglecting the exponentially small terms, after some algebraic transformations, we obtain an expression for $\left.\dfrac{\theta'_\zeta}{\theta}\right|_{\zeta=0}$, the substitution of which into~(\ref{boundary1lindimensionless}) yields 
\begin{equation}
\begin{split}
\dfrac{\varkappa\beta}{\lambda}
\dfrac{X'(\varkappa\beta)}{X(\varkappa\beta)}
\dfrac{
1\!+\!\lambda
\!\left[\!\frac{X(\varkappa\beta)}{\varkappa\beta X'(\varkappa\beta)}
\!+\!\frac{Y'(\varkappa\beta)}{X'(\varkappa\beta)}\!-\!1
\!\right]\!+\!\mu\!\left[\!
\frac{Y(\varkappa|\beta-v|)}{X(\varkappa|\beta-v|)}\!-\!1
\!\right]\!
}
{
1\!+\!\lambda
\left[
\frac{Y(\varkappa\beta)}{X(\varkappa\beta)}\!-\!1
\right]\!+\!\mu\left[
\frac{Y(\varkappa|\beta-v|)}{X(\varkappa|\beta-v|)}\!-\!1
\right]
}
\\
-\nu\dfrac{\varkappa\beta}{\lambda}+
\varepsilon_\text{h}\dfrac{\sigma}{\sigma_m}\left(1-\dfrac{\sigma}{\sigma_m}\right)-
\varepsilon_\text{p}\left(1-2\dfrac{\sigma}{\sigma_m}\right)\!=\!0.
\end{split}\label{thresholdequ}
\end{equation}
This is the threshold equation from which the threshold voltage~$v_\text{th}$ can be found.
In the general case, equation~(\ref{thresholdequ}) admits only a numerical solution.
Note that, in~(\ref{thresholdequ}), for the given voltage~$v$, values of~$\beta$, $\lambda$ and~$\mu$ should be found by minimization of the function~$S(\beta,\lambda,\mu)$~(\ref{S}).
The substitution of the threshold voltage~$v_\text{th}$ into~(\ref{sigma}) yields the respective surface density of adsorbed ions~$\sigma$. 

\begin{figure}[!b]
\centering
\includegraphics[width=60mm]{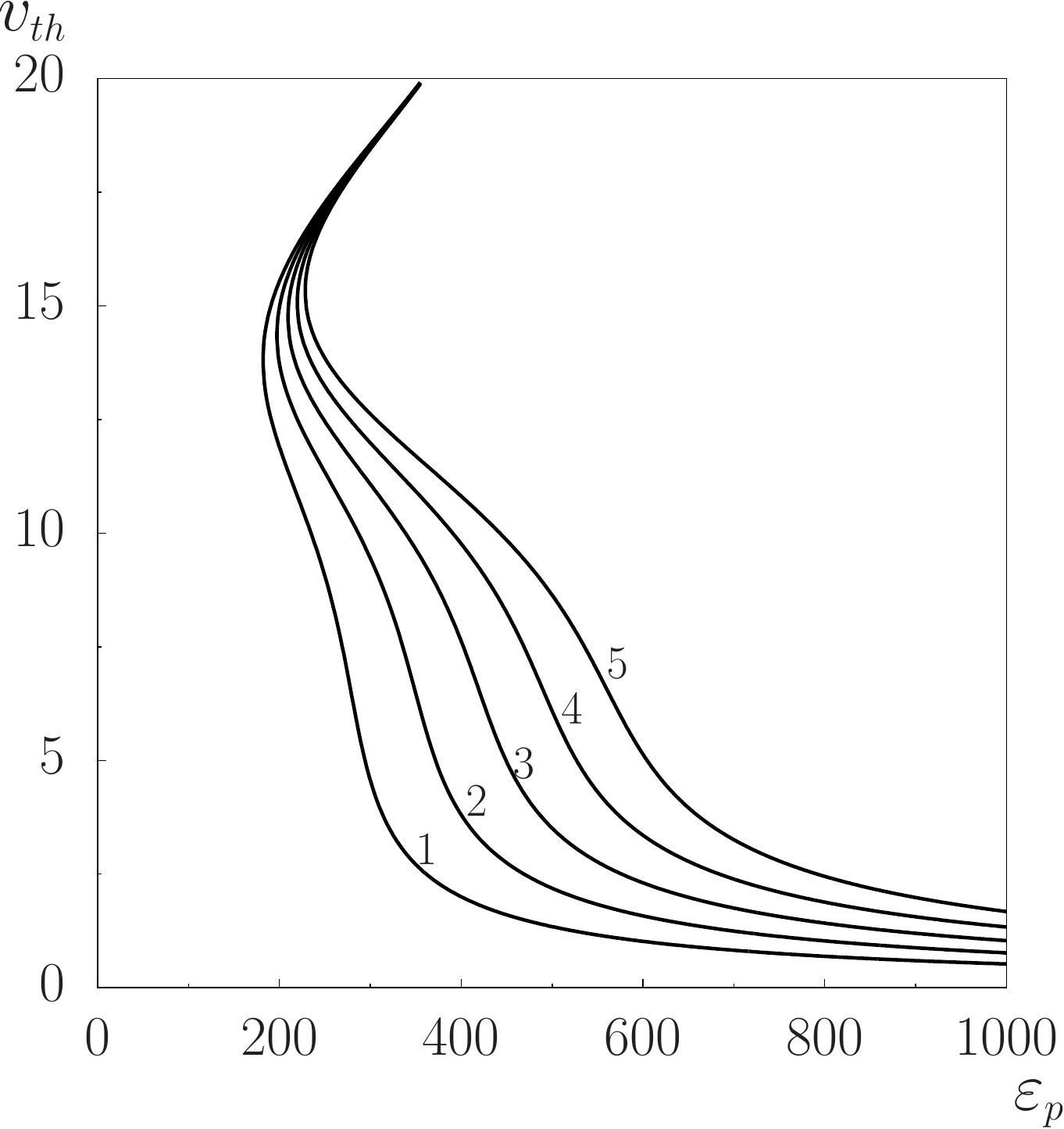}\hspace{15mm}
\includegraphics[width=60mm]{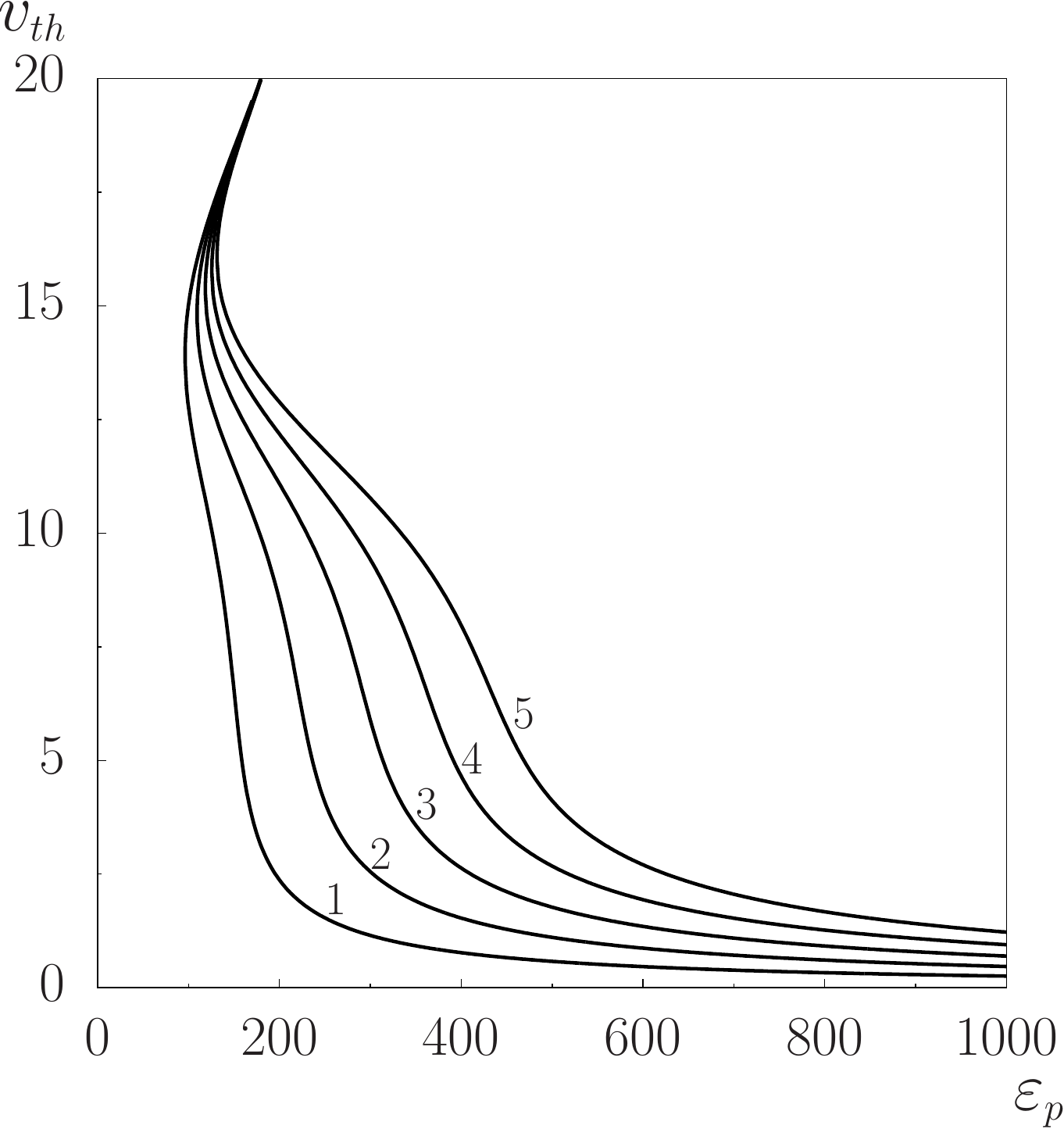}\\[-3mm]
\hspace*{4mm}(a)\hspace{72mm}(b)
\caption{Dimensionless threshold voltage~$v_\text{th}$ as a function of the planar anchoring parameter~$\varepsilon_\text{p}$ in the absence of the flexoelectric polarization~(a) and in the presence of the flexoelectric polarization~($e_1+e_3=20\, \text{pC/m}$)~(b).
$\varepsilon_\text{h} = 0$ (1), $100$ (2), $200$ (3), $300$ (4), $400$ (5).}
\label{vvsep}
\end{figure}

\section{Dependence of threshold voltages on the anchoring parameters}
\label{sec:5}

The dependences of  threshold voltage~$v_\text{th}$ on the planar anchoring parameter~$\varepsilon_\text{p}$, which were obtained by solving equation~(\ref{thresholdequ}), are presented in figure~\ref{vvsep} for the case of the presence of the flexoelectric polarization as well as for the case of its absence. 
Calculations were carried out at several fixed values of the homeotropic anchoring parameter~$\varepsilon_\text{h}$.
We use the values of NLC and CTAB parameters presented in table~\ref{vartable}, which are close to typical~\cite{Sutormin_2012, Sutormin_2014, Sutormin_2016, Barbero_book, Blinov_book}.
\begin{figure}[!t]
\centering
\includegraphics[width=59mm]{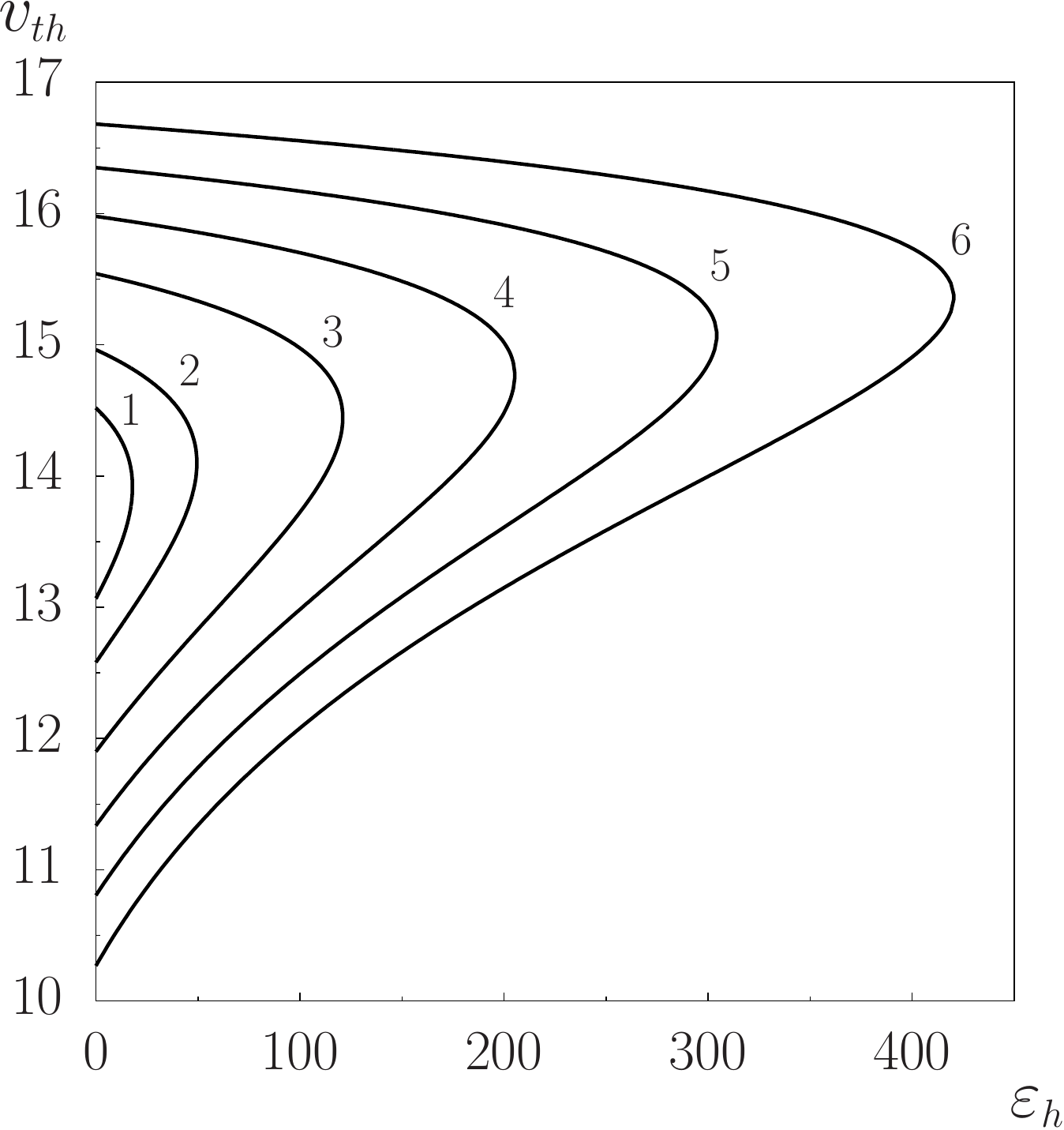}
\hspace{15mm}
\includegraphics[width=60mm]{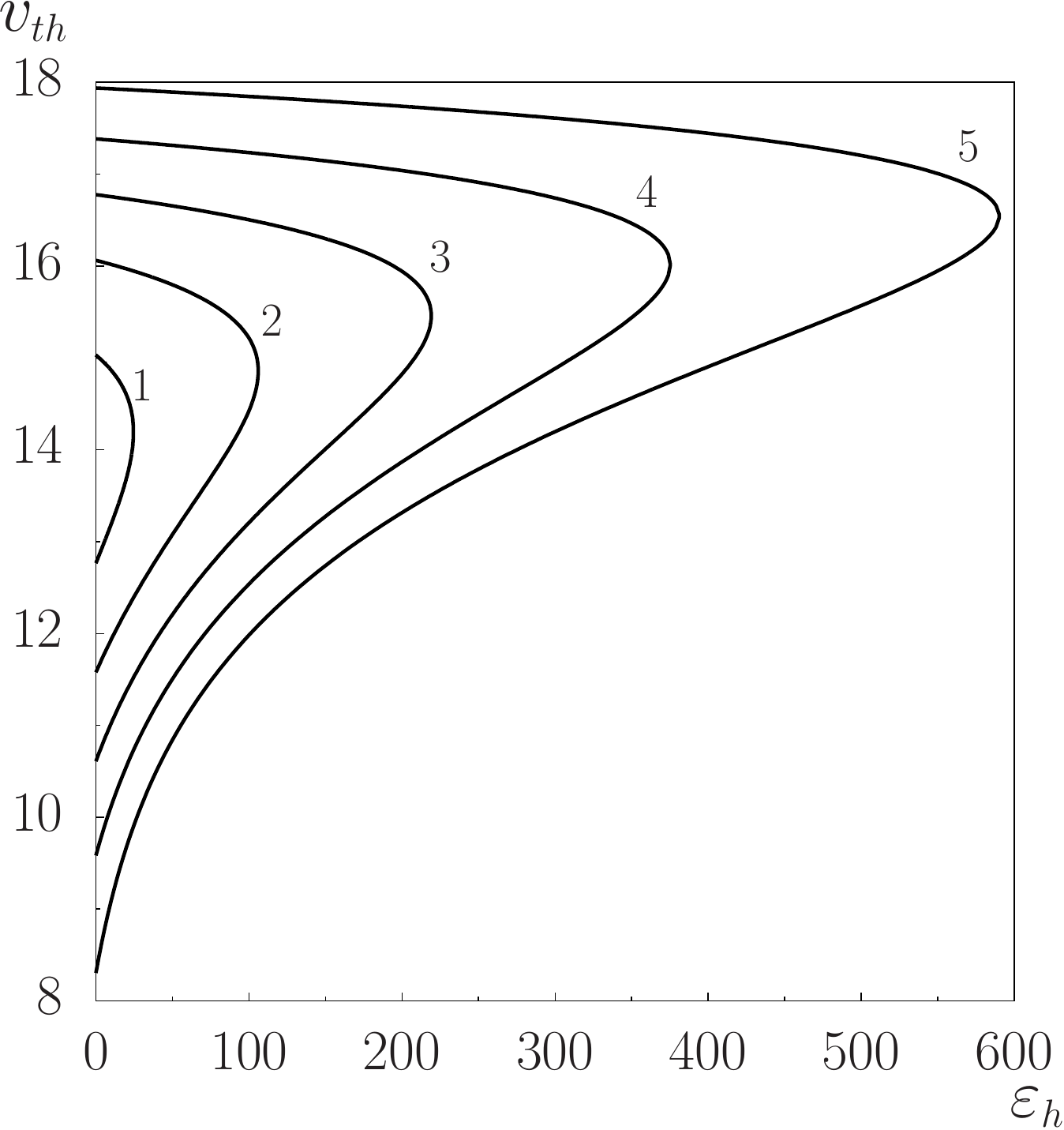}\\[-3mm]
\hspace*{4mm}(a)\hspace{72mm}(b)
\caption{Dimensionless threshold voltage~$v_\text{th}$ versus the homeotropic anchoring parameter~$\varepsilon_\text{h}$ in the absence of the flexoelectric polarization~(a) and in the presence of the flexoelectric polarization ($e_1+e_3=20\, \text{pC/m}$)~(b).
(a)~$\varepsilon_\text{p} = 185$ (1), $190$ (2), $200$ (3), $210$ (4), $220$ (5), $230$ (6);
(b)~$\varepsilon_\text{p} = 100$ (1), $110$ (2), $120$ (3), $130$ (4), $140$ (5).}
\label{vvseh}
\end{figure}

As calculations show, for each given value of~$\varepsilon_\text{h}$
there exists the critical value of the planar anchoring parameter~$\varepsilon_\text{p}^{\text{th}}$.
Hence, at~$\varepsilon_\text{p}<\varepsilon_\text{p}^{\text{th}}$,
the homogeneous homeotropic orientation of the director is preserved in the bulk of the NLC independently of the potential difference~$v$.
Concurrently, at~$\varepsilon_\text{p}>\varepsilon_\text{p}^{\text{th}}$, there are two threshold voltages $v_{\text{th}1}$ and~$v_{\text{th}2}$
($v_{\text{th}1}<v_{\text{th}2}$).
Hence, in voltage ranges~$0\leqslant v\leqslant v_{\text{th}1}$ and~$v\geqslant v_{\text{th}2}$, the homogeneous homeotropic director orientation takes place in the cell.
On the other hand, in the range~$v_{\text{th}1}< v < v_{\text{th}2}$, there is a hybrid homeotropic-planar director orientation in the bulk of the NLC.
The voltage range within which the hybrid director orientation exists broadens, i.e., the threshold voltage~$v_{\text{th}1}$ decreases and ~$v_{\text{th}2}$ increases, with increasing~$\varepsilon_\text{p}$ ($\varepsilon_\text{p}>\varepsilon_\text{p}^{\text{th}}$).
As calculations suggest, the increase in the homeotropic parameter~$\varepsilon_\text{h}$ causes the growth of the critical value~$\varepsilon_\text{p}^{\text{th}}$.

\begin{figure}[!t]
\centering
\includegraphics[width=60mm]{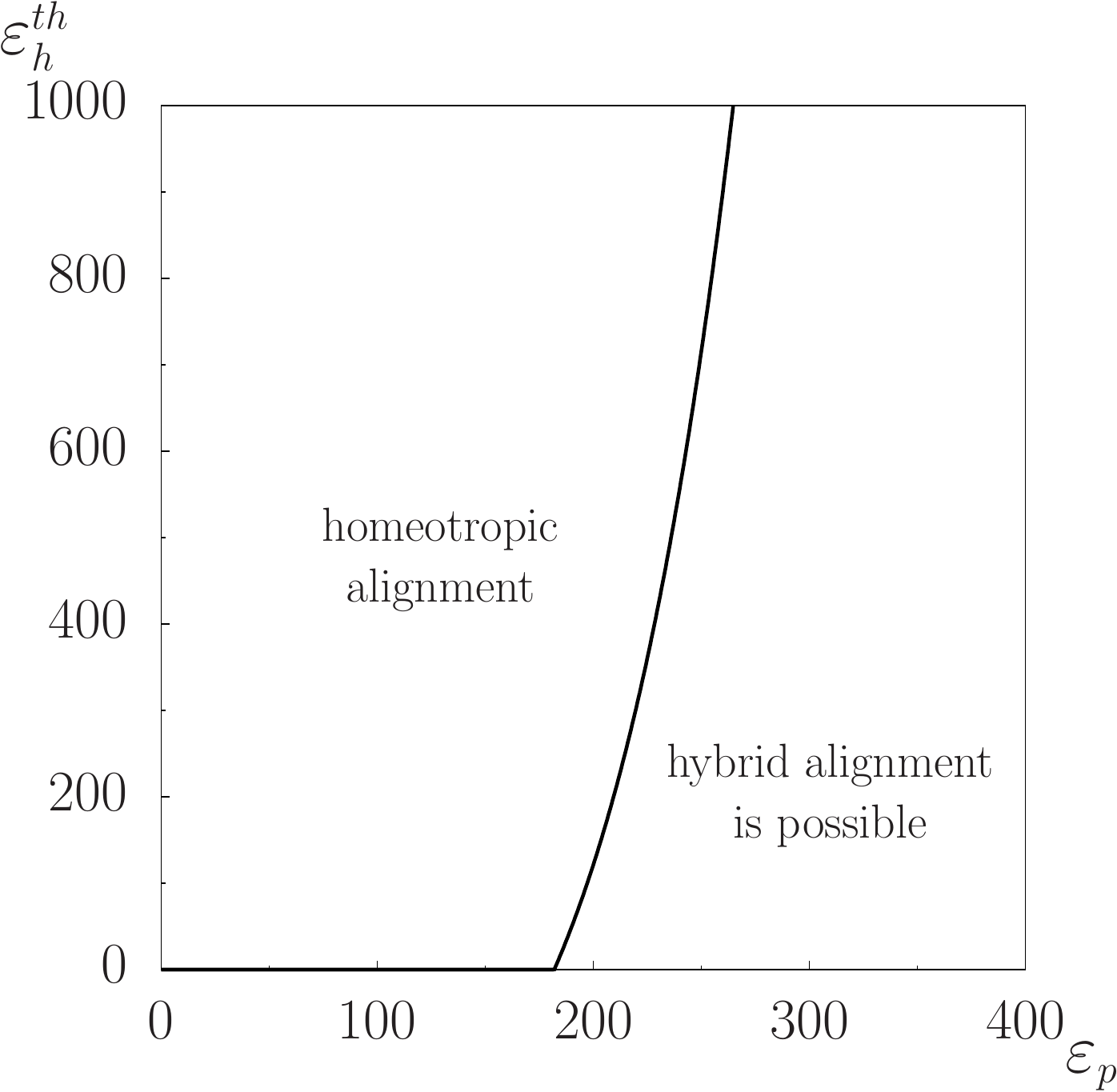}
\hspace{15mm}
\includegraphics[width=60mm]{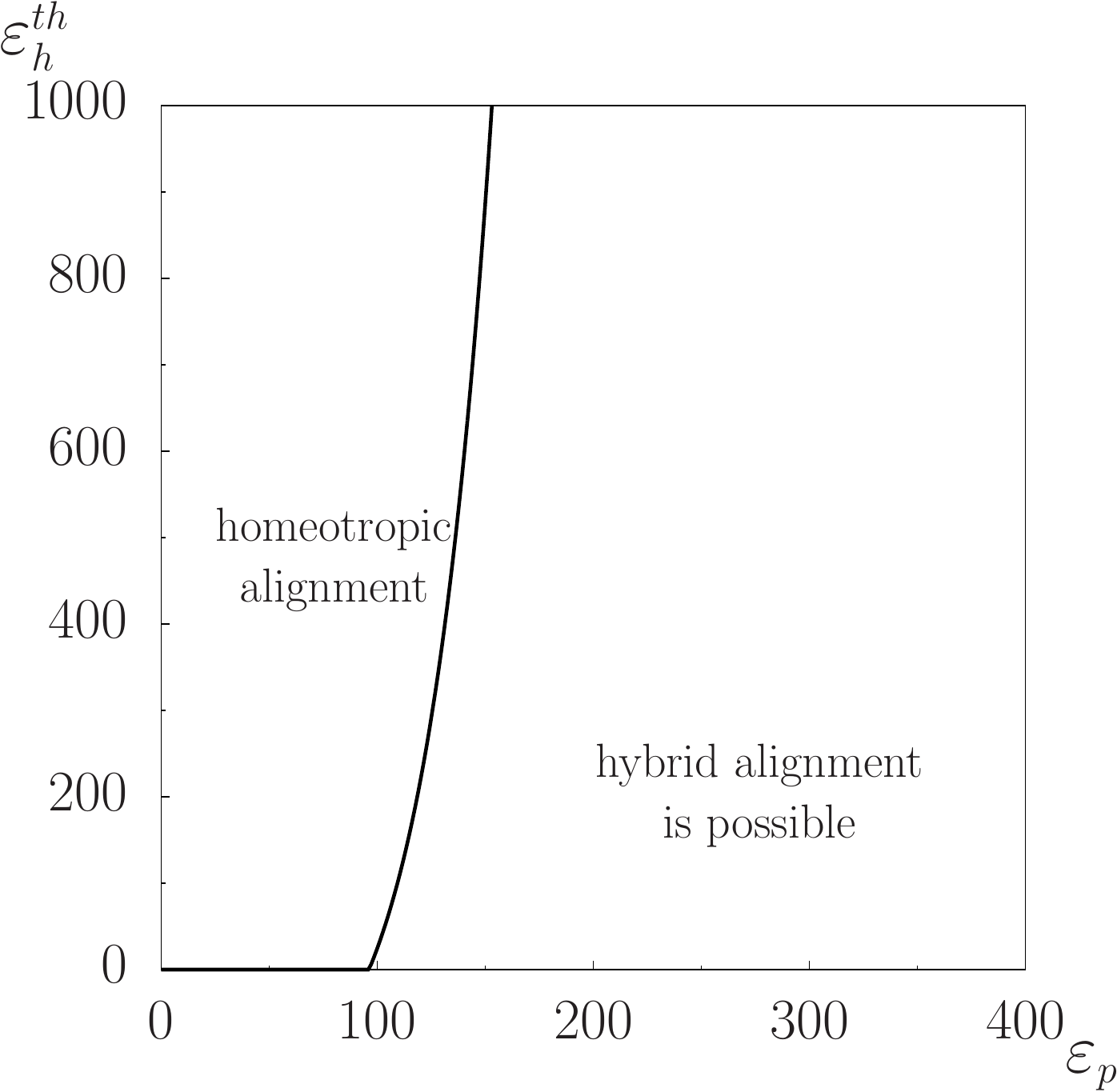}\\[-3mm]
\hspace*{4mm}(a)\hspace{72mm}(b)
\caption{Critical values~$\varepsilon_\text{h}^{\text{th}}$ versus~$\varepsilon_\text{p}$
in the absence of the flexoelectric polarization~(a) and in the presence of the flexoelectric polarization ($e_1+e_3=20\, \text{pC/m}$)~(b).}
\label{epsilonth}
\end{figure}

In figure~\ref{vvseh} the calculated threshold voltage~$v_\text{th}$ as a function of the homeotropic anchoring parameter~$\varepsilon_\text{h}$ is presented in the cases of the presence and the absence of  flexoelectric polarization.
Calculations were carried out at several fixed values of the planar anchoring parameter~$\varepsilon_\text{p}$.
As can be seen, for each value of~$\varepsilon_\text{p}$ there exists some respective critical value of~$\varepsilon_\text{h}^{\text{th}}$.
At~$\varepsilon_\text{h}>\varepsilon_\text{h}^{\text{th}}$, the orientational transition in the bulk of the cell does not take place at any applied voltage~$v$.
The director in the bulk of the NLC preserves the initial homogeneous homeotropic orientation.
However, for each value of~$\varepsilon_\text{h}$ at~$\varepsilon_\text{h}<\varepsilon_\text{h}^{\text{th}}$, there exist respective critical applied voltages~$v_{\text{th}1}$ and~$v_{\text{th}2}$ ($v_{\text{th}1}<v_{\text{th}2}$).
The hybrid homeotropic-planar director orientation takes place only in the voltage range~$v_{\text{th}1}< v < v_{\text{th}2}$.
In the voltage ranges~$0\leqslant v\leqslant v_{\text{th}1}$ and~$v\geqslant v_{\text{th}2}$,
the director remains homogeneous and homeotropically oriented.
The voltage range within which the hybrid homeotropic-planar director orientation exists broadens with decreasing~$\varepsilon_\text{h}$, provided that~$\varepsilon_\text{h}<\varepsilon_\text{h}^{\text{th}}$.

In figure~\ref{epsilonth}, the critical values of~$\varepsilon_\text{h}$ as a function of~$\varepsilon_\text{p}$ are presented for the cases of the absence and the presence of the flexoelectric polarization. 
Hence, at the threshold voltage the orientational transition from the homeotropic director configuration to the hybrid director configuration takes place within the range~$0<\varepsilon_\text{h}<\varepsilon_\text{h}^{\text{th}}$ to which the area to the right of the curve in figure~\ref{epsilonth} corresponds. 
If~$\varepsilon_\text{h}>\varepsilon_\text{h}^{\text{th}}$, then the director in the bulk of the NLC preserves the initial homogeneous homeotropic orientation independently of the applied voltage.  
As can be seen, the range of~$\varepsilon_\text{h}$ within which the threshold orientational instability of the director exists broadens with increasing~$\varepsilon_\text{p}$ independently of the presence of the flexoelectric polarization. 

It is easy to observe that the presence of the flexoelectric polarization does not qualitatively affect the dependence of the threshold voltage~$v_\text{th}$ on homeotropic~$\varepsilon_\text{h}$ and planar~$\varepsilon_\text{p}$ anchoring parameters characterising the anchoring between the NLC and the adsorbing surface.
However, the presence of the flexoelectric polarization expands the voltage range within which the hybrid homeotropic-planar orientation of the director exists. Namely, it causes a decrease in the threshold voltage~$v_{\text{th}1}$ and an increase in~$v_{\text{th}2}$.  
Moreover, the presence of the flexoelectric polarization decreases the critical value~$\varepsilon_\text{p}^{\text{th}}$ of the planar anchoring and increases the critical value~$\varepsilon_\text{h}^{\text{th}}$ of the homeotropic anchoring.

\section{Conclusions}
\label{sec:6}

We have studied the electroinduced director threshold reorientation in a planar flexoelectric NLC cell from the ho\-meo\-tro\-pic configuration to the hybrid homeotropic-planar configuration and vice versa.
The electric field is created by a constant potential difference between the cell surfaces.
The presence of a CTAB-like doping substance in the bulk of the NLC is assumed. A part of the molecules of this substance dissociates in the NLC medium into positive and negative ions.
The orientational instability in the bulk of the NLC is caused by variations of conditions for the NLC director on one of the cell surfaces due to desorption of positive ions on its surface.
The case of a complete screening of the electric field in the bulk is considered, which takes place at relatively small voltages applied to the cell surfaces and large bulk densities of ions~\cite{Sutormin_2012, Sutormin_2014, Sutormin_2016}.
In the frame of the used adsorption model, an approximation to the electric potential profile throughout the cell is obtained. 

It is shown that the presence of ions in the bulk of the NLC and the capability of one of the surfaces to adsorb positive ions considerably affects the threshold voltages. It  also affects the range of homeotropic~$\varepsilon_\text{h}$ and planar~$\varepsilon_\text{p}$ anchoring parameters and the range of the flexoelectric coefficient~$\nu$ within which orientational transitions are possible.  

It is established that for each value of the ho\-meo\-tro\-pic anchoring parameter~$\varepsilon_\text{h}$, there exists the critical value of the planar anchoring parameter~$\varepsilon_\text{p}^{\text{th}}$ so that at~$\varepsilon_\text{p}<\varepsilon_\text{p}^{\text{th}}$ the orientational transition in the bulk of the NLC is impossible independently of the applied voltage~$v$.
Concurrently,  at~$\varepsilon_\text{p}>\varepsilon_\text{p}^{\text{th}}$ there exist two threshold voltages~$v_{\text{th}1}$ and~$v_{\text{th}2}$ (~$v_{\text{th}1}<v_{\text{th}2}$),
so that for the voltages~$v_{\text{th}1}<v<v_{\text{th}2}$ the director configuration is hybrid. 
At~$0\leqslant v \leqslant v_{\text{th}1}$ and~$v\geqslant v_{\text{th}2}$, the director preserves its initial homogeneous homeotropic orientation.
The range of voltages for which the hybrid orientation exists broadens with an increasing~$\varepsilon_\text{p}$. 

It is shown that for each value of~$\varepsilon_\text{p}$, there exists the respective critical value~$\varepsilon_\text{h}^{\text{th}}$ so that
only at~$\varepsilon_\text{h}<\varepsilon_\text{h}^{\text{th}}$ the electroinduced orientational transition in the bulk of the NLC is possible.
If this is the case, the hybrid director configuration emerges at~$v_{\text{th}1}<v<v_{\text{th}2}$, while in the voltage ranges~$0\leqslant v\leqslant v_{\text{th}1}$ and~$v\geqslant v_{\text{th}2}$, the director field preserves the initial homeotropic configuration.
The range of the applied voltage~$v$ within which the hybrid orientation of the NLC director can exist broadens with a decreasing~$\varepsilon_\text{h}$. 

It is important to note that the presence of the flexoelectric polarization expands the ranges of voltage and ho\-meo\-tro\-pic $\varepsilon_\text{h}$ and planar~$\varepsilon_\text{p}$ anchoring parameters within which the hybrid homeotropic-planar NLC director configuration exists. 
\newpage
\section*{Acknowledgements}
The authors are grateful to Prof. Victor Reshetnyak and Prof. Igor Pinkevych for fruitful discussions.

\appendix
\section{Calculation of integrals in the functional $S[u]$~(\ref{functional})}

\setcounter{equation}{0}
\renewcommand\theequation{A.\arabic{equation}}

{ 1.} For the integral $\int\limits_0^1 u'^2_\zeta\rd\zeta$ in the expression for~$S[u]$,
taking into account the explicit form of~$u(\zeta)$~(\ref{ansatz}) and neglecting the exponentially small terms, we obtain
\begin{equation}
\begin{split}
\int\limits_0^1  u'^2_\zeta\;\rd\zeta  &=\dfrac{\beta^2}{\lambda^2}\int\limits_0^1 \re^{-\frac{2\zeta}{\lambda}}\;\rd\zeta+
\dfrac{2\beta(v-\beta)}{\lambda\mu}\re^{-\frac{1}{\mu}}\int\limits_0^1 \re^{\left(\frac{1}{\mu}-\frac{1}{\lambda}\right)\zeta}\;\rd\zeta
  \\ & +
\dfrac{(v-\beta)^2}{\mu^2}\int\limits_0^1 \re^{-\frac{2(1-\zeta)}{\mu}}\;\rd\zeta
=\dfrac{\beta^2}{2\lambda}\bigl(1-\re^{-\frac{2}{\lambda}}\bigr)
 \\ & +
2\beta(v-\beta)\dfrac{\re^{-\frac{1}{\lambda}}-\re^{-\frac{1}{\mu}}}{\lambda-\mu}+
\dfrac{(v-\beta)^2}{2\mu}\bigl(1-\re^{-\frac{2}{\mu}}\bigr)
\approx \dfrac{\beta^2}{2\lambda} + \dfrac{(v-\beta)^2}{2\mu}. \label{I1}
\end{split}
\end{equation}

\noindent
{ 2.} The second integral $\int\limits_0^1 \re^{-u(\zeta)}\;\rd\zeta$ in the expression for~$S[u]$
can be calculated as follows
\begin{equation}
\begin{split}
 \int\limits_0^1 \re^{-u(\zeta)}\;\rd\zeta & =
\int\limits_0^1 \re^{-\beta \re^{-\frac{\zeta}{\lambda}}}\re^{-(v-\beta)\re^{-\frac{1-\zeta}{\mu}}}\re^\beta\;\rd\zeta 
\\ \!&
 =
\re^\beta \int\limits_0^1\!\left(\!\left[\re^{-\beta \re^{-\frac{\zeta}{\lambda}}}-1\right]\!+\!1\right)\!
\left(\left[\re^{-(v-\beta)\re^{-\frac{1-\zeta}{\mu}}}-1\right]\!+\!1\right)\rd\zeta.
\end{split}
\end{equation}
Here, the expressions in square brackets are substantially different from~$0$ only in the vicinities of the points~$\zeta=0$ and~$\zeta=1$, respectively.
Therefore, with the accuracy up to exponentially small terms, one can write

\begin{equation*}
\int\limits_0^1 \re^{-u(\zeta)}\;\rd\zeta =\re^\beta \!\left(\!\int\limits_0^1 \re^{-\beta \re^{-\frac{\zeta}{\lambda}}}
\rd\zeta\!+\!\int\limits_0^1 \re^{-(v-\beta)\re^{-\frac{1-\zeta}{\mu}}}\rd\zeta\!-\!1\right)\!.
\end{equation*}

Using the substitution~$\xi=1-\zeta$ in the second integral, we have 
\begin{equation}\label{I2}
\begin{split}
\int\limits_0^1 \re^{-u(\zeta)} \;\rd\zeta  &
=\re^\beta \!\left(\!\int\limits_0^1 \re^{-\beta \re^{-\frac{\zeta}{\lambda}}}
\rd\zeta+\int\limits_0^1 \re^{-(v-\beta)\re^{-\frac{\xi}{\mu}}}\rd\xi\!-\!1\!\right)\!  \\
& \equiv \re^\beta \left(I(-\beta,\lambda)+I(v-\beta,\mu)-1\right),
\end{split}
\end{equation}
where the integral
$\displaystyle
I(\beta,\lambda)=\int\limits_0^1 \re^{\beta \re^{-\zeta/\lambda}}\;\rd\zeta
$
is calculated below (see paragraph 3).

Similarly, the third integral in expression~(\ref{functional}) equals
\begin{equation}\label{I3}
\int\limits_0^1 \re^{u(\zeta)}\;\rd\zeta= \re^{-\beta} \bigl[I(\beta,\lambda)+I(\beta-v,\mu)-1\bigr].
\end{equation}

\noindent
{3.} Using the substitution $t=\beta \re^{-\frac{\zeta}{\lambda}}$, the integral
$\displaystyle
I(\beta,\lambda)=\int\limits_0^1 \re^{\beta \re^{-\zeta/\lambda}}\;\rd\zeta
$ can be transformed into
\begin{equation}\label{Ires}
\begin{split}
\! & I(\beta,\lambda)
=\lambda\int\limits_{\beta \re^{-\frac{1}{\lambda}}}^{\beta}\dfrac{\re^t}{t}\;\rd t=
\lambda\int\limits_{\beta \re^{-\frac{1}{\lambda}}}^{\beta}\dfrac{\rd t}{t}+
\lambda\int\limits_{\beta \re^{-\frac{1}{\lambda}}}^{\beta}\dfrac{\re^t-1}{t}\;\rd t \\
& =
\Big.\lambda {\rm ln}|t|\Big|_{\beta \re^{-\frac{1}{\lambda}}}\!+\!
\lambda\!\!\!\int\limits_{\beta \re^{-\frac{1}{\lambda}}}^{\beta}\!\!\!\dfrac{\sum\limits_{n=1}^{\infty}\frac{t^n}{n!}}{t}\rd t\!=\!
1\!+\!\lambda\!\left[\sum\limits_{n=1}^{\infty}\dfrac{t^n}{n\cdot n!}\right]_{t=\beta \re^{-\frac{1}{\lambda}}}^{t=\beta}\!\!\!\!\!  \\
& = 1+\lambda f(\beta)-\lambda f(\beta \re^{-\frac{1}{\lambda}})\approx 1+\lambda f(\beta),
\end{split}
\end{equation}
where~$f(\beta)=\sum\limits_{n=1}^{\infty}\dfrac{\beta^n}{n\cdot n!}$.

We next substitute $I(\beta,\lambda)$~(\ref{Ires}) into~(\ref{I2}) and~(\ref{I3}). 
Taking into account~(\ref{I1}), (\ref{I2}) and~(\ref{I3}), we arrive at
the final form of the functional $S(\beta,\lambda,\mu)$~(\ref{S}).

\ukrainianpart

\title{Орієнтаційна нестійкість директора в нематичній комірці спричинена
електроіндукованою зміною зчеплення}
\author{О.С. Тарнавський, М.Ф. Ледней}
\address{
Фізичний факультет, Київський національний університет імені Тараса Шевченка,
проспект Академіка Глушкова 4, 03022, Київ, Україна
}
\makeukrtitle

\begin{abstract}
\tolerance=3000%
Теоретично досліджено порогову переорієнтацію директора із гомеотропного стану в гібридний гомеотроп-планарний
і навпаки у плоскопаралельній комірці флексоелектричного
нематичного рідкого кристалу (НРК) в електричному полі.
Рідкий кристал доповано речовиною типу~СТАВ, частина молекул якої дисоціює на позитивно і негативно заряджені іони.
Зчеплення~НРК з однією із підкладок комірки є гомеотропним. На поверхні іншої підкладки можлива адсорбція позитивно заряджених іонів, які відіграють роль орієнтанта молекул~НРК на цій поверхні. 
При певних величинах прикладеної напруги можливий орієнтаційний перехід директора в об'ємі~НРК з гомеотропного стану в гібридний і/або навпаки.
Розраховані величини порогових напруг орієнтаційних переходів у залежності від значень параметрів зчеплення~НРК з поверхнею адсорбуючої підкладки. 
Встановлено існування критичних значень параметрів зчеплення, що визначають області зміни цих параметрів в межах яких мають місце орієнтаційні переходи.

\keywords нематичний рідкий кристал, орієнтаційна нестійкість, поріг орієнтаційної нестійкості, межові умови, флексополяризація

\end{abstract}

\end{document}